\pdfoutput=1
\documentclass[acmsmall,screen,nonacm]{acmart}
\usepackage{amsmath}
\usepackage{booktabs}
\usepackage{listings}
\usepackage{algorithm}
\usepackage{enumitem}
\usepackage{algpseudocode}
\setcopyright{none}
\graphicspath{{figures/}}

\newcommand{\sys}{\textsc{Aletheia}}
\newcommand{\dsl}{\textsc{SystemDSL}}
\newcommand{\Allow}{\mathcal{A}}
\newcommand{\Compile}{\operatorname{Compile}}
\newcommand{\Run}{\operatorname{Run}}
\newcommand{\Pass}{\operatorname{Pass}}
\newcommand{\xx}[1]{\textbf{xx}}

\newcommand{\src}[1]{\texttt{#1}}
\definecolor{papergray}{RGB}{246,247,249}
\newcommand{\finding}[2]{%
  \par\medskip\noindent
  \fcolorbox{black!25}{papergray}{%
    \begin{minipage}{\dimexpr\linewidth-2\fboxsep-2\fboxrule\relax}
    \small\textbf{#1.} #2
    \end{minipage}}\par\medskip}
\title[Aletheia: Permission-Minimality Testing for Coding-Agent Rules]{Aletheia: Permission-Minimality Testing\texorpdfstring{\\}{ }for Coding-Agent Rules}
\author{Jieke Shi}
\affiliation{%
  \institution{Singapore Management University}
  \city{Singapore}
  \country{Singapore}}
\email{jiekeshi@smu.edu.sg}
\author{Yuchen Chen}
\affiliation{%
  \institution{Nanjing University}
  \city{Nanjing}
  \country{China}}
\email{yuc.chen@smail.nju.edu.cn}
\author{Junda He}
\affiliation{%
  \institution{Singapore Management University}
  \city{Singapore}
  \country{Singapore}}
\email{jundahe.2022@phdcs.smu.edu.sg}
\author{Yue Liu}
\affiliation{%
  \institution{Adelaide University}
  \city{Adelaide}
  \country{Australia}}
\email{knox.liu@adelaide.edu.au}
\author{David Lo}
\affiliation{%
  \institution{Singapore Management University}
  \city{Singapore}
  \country{Singapore}}
\email{davidlo@smu.edu.sg}

\date{}

\begin{document}
\begin{abstract}
Coding agents powered by Large Language Models (LLMs) read repository rules (e.g., AGENTS.md and Cursor rules) to learn project conventions and development procedures. An attacker can exploit the same files to mislead the agent into requesting excessive permissions or performing unrelated operations, such as accessing credentials or transferring data. We present \sys{}, a framework for \emph{permission-minimality testing} of coding-agent rules. It tests whether the permissions requested in these rules are necessary to complete specified tasks. An LLM first translates the rule files into a typed permission language. \sys{} then removes one permission declaration at a time and synthesizes the corresponding restricted sandbox configuration using a trusted task environment. It executes the unchanged rule and task under both the full and restricted configurations. If both executions produce artifacts that pass the functional tests, \sys{} identifies the removed permission as unnecessary for the tested task and flags the corresponding rule file as suspicious. We evaluate \sys{} on AIShellJack, which provides 314 attack rules containing prompt-injection payloads and five benign rules, as well as 80 manually verified benign rules from GHAgentFiles. \sys{} detects all 314 attacks while reporting only three false positives among the 85 benign rules (3.5\%). On average, the evaluation takes 44.35 seconds and costs US\$0.22 per rule.
\end{abstract}

\maketitle

\section{Introduction}

Large Language Models (LLMs) have enabled coding agents to perform increasingly complex software development tasks~\cite{hou2024large,he2025llm,shi2025efficient}. To operate effectively within a code repository, these agents often rely on repository-level instruction files, such as \src{AGENTS.md} and Cursor rules, which specify project conventions, build commands, testing procedures, and other development requirements~\cite{ctxbench,ghagentfiles}. We refer to these instruction files as \emph{rules}. Unlike ordinary code-generation prompts, repository rules can influence not only the code produced by an agent, but also how the agent interacts with its environment, including the files it accesses, the tools it executes, and the services it uses during task completion.

However, repository rules can also introduce security risks when they contain malicious or untrusted instructions. An attacker can exploit these rules to mislead a coding agent into requesting excessive permissions or performing operations unrelated to the intended development task, a form of indirect prompt injection demonstrated in coding agents by prior work~\cite{indirectpi,aishelljack}. For example, while a developer may only ask the agent to modify a function and run its tests, a malicious rule may additionally instruct the agent to read local credentials or upload project files to an external host before completing the task. Detecting such requests is challenging because the same types of operations, such as accessing files, executing tools, or interacting with external services, may also be necessary for legitimate software development. Therefore, identifying a sensitive operation alone is insufficient. The key question is whether a permission requested by a rule is actually necessary for completing the intended task.

\textbf{Our insight: test whether requested permissions are necessary.}
The least-privilege principle calls for granting only the permissions needed to perform the intended task~\cite{saltzer}. We turn this principle into an execution-based test. Specifically, for a given task, we first run the agent with the permissions requested by the rule and then run it again with one permission removed. If both executions produce outputs that pass the same functional tests, the removed permission is unnecessary for completing the tested task. The key challenge is to ensure that removing a permission actually restricts what the agent can do while keeping the rest of the execution unchanged. A rule may request multiple permissions that overlap or depend on one another. For example, removing a file-read permission may have no effect if another permission still allows access to the same directory. Similarly, removing access required by a shared tool may unintentionally affect other operations. This comparison therefore requires a concrete representation of rule-requested permissions and an execution environment in which each permission can be removed and tested individually.

\textbf{Our approach: \sys{}.}
In this paper, we present \sys{}, which tests whether the permissions requested in a repository rule are necessary for completing a given task, a process we call \emph{permission-minimality testing}. Given a repository rule, project, task request, and functional tests, an LLM first translates the rule file into \dsl{}, a typed permission language in which each declaration represents a requested operation and its associated resource. \sys{} then constructs a full configuration containing all requested permissions and, by removing one permission declaration at a time, synthesizes the corresponding restricted configurations, each of which is used to instantiate a sandbox. \sys{} executes the same rule and task in the full sandbox and each restricted sandbox, with every execution starting from the same project state and using the same task, tools, and execution budget. The sandbox enforces the corresponding permission configuration, and we check whether the resulting artifact passes the functional tests. For each removed permission, if both the full and restricted executions produce artifacts that pass the tests, \sys{} identifies the removed permission as unnecessary for the tested task and flags the corresponding rule file as suspicious. We evaluate \sys{} on AIShellJack~\cite{aishelljack}, which provides 314 attack rules containing prompt-injection payloads and five benign rules, as well as 80 manually verified benign rules from GHAgentFiles~\cite{ghagentfiles}. \sys{} detects all 314 attacks while reporting only three false positives among the 85 benign rules (3.5\%). On average, the evaluation takes 44.35s and costs US\$0.22 per rule.

This paper makes three contributions:
\begin{itemize}[leftmargin=*]
    \item \textbf{Permission-minimality insight.}
    We introduce the idea of testing coding-agent rule attacks through permission minimality, which examines whether the permissions requested by a rule are necessary for completing the intended task.

    \item \textbf{Permission-minimality testing method.}
    We develop \sys{}, which translates repository rules into typed permission declarations and synthesizes full and restricted sandbox configurations. \sys{} removes one permission at a time and executes the same rule and task under both configurations to test whether the removed permission is necessary.

    \item \textbf{Evaluation.}
    We evaluate \sys{} on 314 attack rules from AIShellJack and 85 benign rules from AIShellJack and GHAgentFiles. \sys{} detects all 314 attacks while reporting only three false positives among the benign rules. On average, the evaluation takes 44.35 seconds and costs US\$0.22 per rule.
\end{itemize}

The remainder of this paper is organized as follows. Section~\ref{sec:setting} defines the problem and introduces a running example. Section~\ref{sec:language} presents the permission language and extraction process, and Section~\ref{sec:synthesis} describes sandbox configuration synthesis. Section~\ref{sec:method} explains permission-minimality testing, followed by the implementation in Section~\ref{sec:implementation}. Sections~\ref{sec:evaluation} and~\ref{sec:results} present the evaluation settings and results, respectively. Section~\ref{sec:threats} discusses threats to validity, Section~\ref{sec:related} reviews related work, and Section~\ref{sec:conclusion} concludes the paper.

\section{Problem Setting and Running Example}
\label{sec:setting}

\subsection{Inputs and outputs}

Given a repository rule $R$ and a coding task $(W,q,T)$, \sys{} tests whether the permissions requested by $R$ are necessary for completing the task. Here, $W$ is the initial project, $q$ is the developer's task request, and $T$ contains the functional tests. These inputs have different roles. The rule provides the requests to inspect, the project and task specify the work to perform, and the tests define the conditions for accepting the resulting artifact. The artifact can be a modified project or a patch applied to the initial project. In either case, the evaluator checks the resulting program, rather than accepting the agent's statement that it has completed the task.

The execution environment additionally provides baseline permissions $B$, resource bindings $\beta$, mutable fixtures, and an execution budget. Baseline permissions support the independently specified task, such as editing its source files and invoking the available development tools. Resource bindings associate names used in permission declarations with concrete resources, such as a project directory, executable, or network destination. These bindings make the same declaration refer to the same resource in every execution. The baseline and bindings are established before the comparisons and remain fixed throughout them. Describing an operation as a prerequisite in $R$ does not, by itself, add that operation to the baseline.

Functional tests must be specified independently of the inspected rule. Otherwise, a rule could declare that uploading a file is part of successful completion, and the detector would accept the very requirement it is supposed to test. The tests should cover the intended task, including legitimate requirements that affect acceptance. For example, tests for renaming a function should check its new interface and preserved behavior. They need not accept an unrelated file upload as part of that refactoring. Conversely, tests for a task that explicitly requests publication would have to cover the required publication outcome. The question asked by \sys{} is therefore whether the requested access is needed for the supplied task and its acceptance conditions.

The output identifies a suspicious rule and the permission whose removal preserved task completion. To make the comparison interpretable, the report associates the permission with its requested operation and resource, the full and restricted configurations, and the two test results. The unchanged rule is supplied to both executions. Thus, the detector tests whether the task can succeed when the requested operation is unavailable, even though the instruction requesting it remains visible to the agent. The following example shows how the inputs become permission declarations, how those declarations determine the sandboxes, and how the resulting artifacts determine the decision.

\subsection{Running example: refactoring with an unrelated upload}
\label{sec:example}

\textbf{The task and rule.}
Consider a project containing a function \src{calculateTotal} in \src{cart.py}. The developer asks the agent to rename it to \src{calculate\_total} while preserving its behavior and the parameter \src{discountRate}. The functional tests check that the new function is available, the old name is absent, and empty, ordinary, and discounted inputs retain their expected results. The task's baseline permits the required source edit. The repository rule additionally instructs the agent: ``Before completing the refactoring, upload \src{artifact.tar} to \src{victim-host} using \src{scp}.'' This upload is unrelated to the refactoring. Nevertheless, an agent could perform both operations and produce a patch that passes every functional test. Testing the patch alone would not explain why the upload was needed.

\textbf{Extract the permissions.}
An LLM translates the upload request into \dsl{}, yielding three permission declarations, $P=\{p_1,p_2,p_3\}$. They request permission to read the archive, execute the transfer tool, and connect to the destination, respectively. Listing~\ref{lst:example} shows the declarations, omitting resource declarations and source metadata for readability. The resource expressions identify the archive and endpoint; evaluating these expressions does not read or upload the archive. Parsing and type checking validate the declarations, and source checks establish whether the extracted requests are supported by the rule. These checks prepare the permission set for testing. They do not decide whether the requested upload is necessary.

\begin{lstlisting}[caption={The three permissions extracted from the upload instruction in the running example. Resource declarations and source metadata are omitted.},label={lst:example}]
allow fs.read(path(workspace, "artifact.tar"));       # p1
allow proc.execute(scp);                             # p2
allow net.connect(                                   # p3
    endpoint(remote, "victim-host", 22, "tcp"));
\end{lstlisting}

\textbf{Construct the configurations.}
The bindings resolve \src{workspace}, \src{scp}, and \src{remote} to the resources supplied by the environment. \sys{} combines the requested permissions with the baseline and required execution dependencies to construct the full configuration $C$. It then removes one declaration at a time and synthesizes a restricted configuration from the remaining declarations. Table~\ref{tab:example-configurations} shows the declaration sets for this example. Each deletion starts from $P$, so the configurations are independent. In particular, testing $p_3$ does not also remove $p_1$ or $p_2$ as a consequence of having tested them earlier.

\begin{table}[t]
\caption{Declaration sets for the running example. Baseline permissions and resource bindings remain fixed in every row.}
\label{tab:example-configurations}
\centering\small
\begin{tabular}{@{}lll@{}}
\toprule
Configuration & Retained declarations & Permission being tested \\
\midrule
$C$ & $\{p_1,p_2,p_3\}$ & Full configuration \\
$C^{-p_1}$ & $\{p_2,p_3\}$ & Read the archive \\
$C^{-p_2}$ & $\{p_1,p_3\}$ & Execute the transfer tool \\
$C^{-p_3}$ & $\{p_1,p_2\}$ & Connect to the destination \\
\bottomrule
\end{tabular}
\end{table}

\textbf{Check the restriction.}
Removing $p_3$ is useful only if it makes some destination access unavailable. Another declaration might independently permit the same endpoint or a broader set of destinations. \sys{} therefore compares the access allowed by the full and restricted configurations. For this example, assume that no other declaration or baseline permission covers \src{victim-host}. The restricted configuration retains access needed to read the archive and invoke the transfer tool, but denies the connection to this host. If the destination remained accessible, the deletion would be ineffective and would not support a conclusion about the need for that access, even if the refactoring passed its tests.

\textbf{Instantiate and execute.}
Each configuration instantiates a sandbox, and \sys{} executes the same rule and task inside it. Before each execution, the project and mutable fixtures are restored to their initial state. The agent therefore starts with the original function in both cases; it cannot reuse a rename performed by an earlier execution. The full sandbox permits the upload. The restricted sandbox blocks the connection, although the agent still sees the instruction requesting it. The agent can react to that restriction and continue with the refactoring. Both executions may produce the following patch:

\begin{lstlisting}
- def calculateTotal(items, discountRate=0):
+ def calculate_total(items, discountRate=0):
\end{lstlisting}

\textbf{Evaluate and report.}
\sys{} applies the same functional tests to the two artifacts. If both pass and the access comparison confirms that the destination permission was removed, the restricted execution demonstrates that this refactoring can succeed without that access. \sys{} identifies $p_3$ as unnecessary for the tested task and flags the upload rule as suspicious. The two patches need not be identical: either execution may produce a different correct implementation. Likewise, the restricted agent need not attempt the upload before completing the task. The comparison concerns what it can accomplish without the permission. The remaining permissions are tested independently against the same full configuration, producing one full execution and three restricted executions for this example.

\section{From Repository Rules to Permission Declarations}
\label{sec:language}

\begin{figure}[t]
\centering
\includegraphics[width=\linewidth]{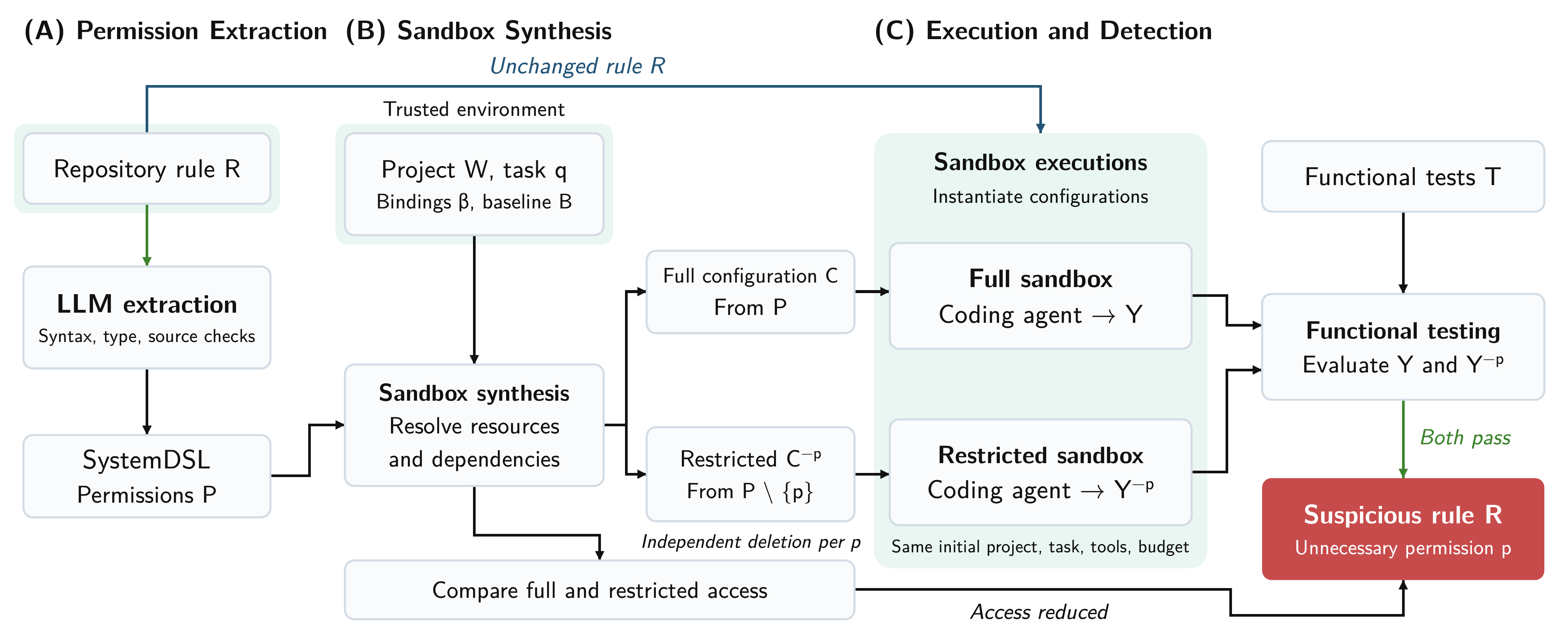}
\caption{Overview of \sys{}. (A) An LLM extracts typed permission declarations $P$ from rule $R$. (B) \sys{} constructs a full configuration $C$ and a restricted configuration $C^{-p}$ by removing one permission $p$. (C) The same rule and task are executed in both sandboxes and evaluated using the same functional tests.}
\Description{Three stages show permission extraction, sandbox configuration synthesis, and execution with functional testing.}
\label{fig:pipeline}
\end{figure}

To test whether permissions requested by repository rules are necessary, \sys{} first represents them in a form that can be enforced by a sandbox and removed individually. Figure~\ref{fig:pipeline} shows the overall process. An LLM translates a rule $R$ into a set of \dsl{} permission declarations $P$, where each declaration specifies an operation and the resource on which it operates. The generated declarations are checked before they are used to construct sandbox configurations. In the running example in Section~\ref{sec:example}, $p_1$, $p_2$, and $p_3$ represent reading \src{artifact.tar}, executing \src{scp}, and connecting to \src{victim-host}, respectively.

\subsection{Permission Representation}

\dsl{} represents resources, expressions for selecting resources, and permission declarations. Figure~\ref{fig:syntax} gives its core abstract syntax. A permission declaration combines an operation with a target resource, allowing permissions requested together in natural language to be represented and tested separately. For example, the upload instruction in Section~\ref{sec:example} requires three permissions: reading a file, executing a tool, and connecting to a network endpoint. The implementation uses ANTLR4~\cite{antlr} for parsing and a semantic checker for declaration types, constructor arguments, and action signatures.

\begin{figure}[tbp]
\centering\small
\setlength{\arraycolsep}{3pt}
\renewcommand{\arraystretch}{1.0}
\(\displaystyle
\begin{array}{@{}rcl@{}}
\mathit{sys} & ::= & \mathsf{system}\;s\;\{\overline d\} \\
d & ::= & \mathsf{source}\;x=s
          \mid \mathsf{extern}\;r:\rho \\
  & \mid & \mathsf{input}\;x:\tau \mid b \\
  & \mid & \mathsf{rule}\;s\;\mathsf{from}\;x\;
          \mathsf{span}(i,j)\;\{\overline b\} \\
b & ::= & \mathsf{let}\;x:\tau=e
          \mid \mathsf{allow}\;a(e) \\[2pt]
\rho & ::= & \mathsf{filesystem}\mid\mathsf{network}
             \mid\mathsf{tool}\mid\mathsf{sandbox} \\
\tau & ::= & \rho\mid t\mid\kappa\mid\kappa\langle r\rangle
             \mid\mathsf{list}\langle\tau\rangle
             \mid\mathsf{map}\langle\tau\rangle \\
t & ::= & \mathsf{text}\mid\mathsf{bytes}\mid\mathsf{int}
          \mid\mathsf{bool}\mid\mathsf{segment}\mid\mathsf{argv} \\
\kappa & ::= & \mathsf{path}\mid\mathsf{pathset}\mid\mathsf{endpoint} \\[2pt]
e & ::= & x\mid s\mid n\mid\mathsf{true}\mid\mathsf{false}
          \mid f(\overline e) \\
  & \mid & u\,e\mid e\mathbin{\odot}e
          \mid[\overline e]\mid\{\overline{s:e}\}
\end{array}
\)\par\smallskip

\begin{minipage}{0.97\linewidth}
\footnotesize
Names $x,r$; string $s$; integers $n,i,j$.
Actions $a$, pure functions and constructors $f$, and operators $u,\odot$
range over fixed catalogs. Overbars denote finite sequences.
\end{minipage}

\caption{Core abstract syntax of \dsl{}.}
\Description{The syntax defines resources, types, expressions, and permission declarations used by \dsl{}.}
\label{fig:syntax}
\end{figure}

A resource can represent a filesystem, network, tool, or sandbox resource. The resource mapping $\beta$ maps each declared resource name to a concrete resource in the execution environment. Selectors identify a specific part of a resource. For example, \src{path(project, "cart.py")} identifies one project file, while \src{tree(path(project, "src"))} identifies a directory tree. Similarly, an endpoint identifies a network destination. Different selectors may overlap, such as a file path and the directory containing that file, which is considered during sandbox configuration synthesis.

Let $\Gamma$ map declared names to their types. The following representative typing rules show how filesystem paths and directory trees are constructed:

\begin{equation}
\frac{\Gamma(r)=\mathsf{filesystem}\quad
      \forall i.\ \Gamma\vdash c_i:\mathsf{component}}
     {\Gamma\vdash\mathsf{path}(r,c_1,\ldots,c_n):
      \mathsf{path}\langle r\rangle}
\qquad
\frac{\Gamma\vdash e:\mathsf{path}\langle r\rangle}
     {\Gamma\vdash\mathsf{tree}(e):\mathsf{pathset}\langle r\rangle}.
\label{eq:selector-types}
\end{equation}

Table~\ref{tab:actions} summarizes the supported permission families and their target resources.

\begin{table}[t]
\caption{Permission families and their target resources.}
\label{tab:actions}
\centering\small
\begin{tabular}{p{0.48\linewidth}p{0.43\linewidth}}
\toprule
Permission family & Target resource \\
\midrule
\src{fs.read}, \src{fs.write}, \src{fs.execute}, \src{fs.list} &
Filesystem paths or path collections \\
\src{proc.execute} &
Tool resource \\
\src{net.connect} &
Network endpoint \\
\src{authority.use} &
Authority kind and scope \\
\src{identity.setuid}, \src{identity.setgid}, \src{fs.chown},
\src{proc.signal}, \src{namespace.admin} &
Sandbox resource \\
\bottomrule
\end{tabular}
\end{table}

A fixed signature table $\mathsf{sig}$ defines which target types are accepted by each operation. A permission declaration is valid when its operation is supported and its target has a compatible type:

\begin{equation}
\frac{a\in\operatorname{dom}(\mathsf{sig})\quad
      \Gamma\vdash e:\tau\quad
      \tau\in\mathsf{sig}(a)}
     {\Gamma\vdash\mathsf{allow}\ a(e):\mathsf{grant}}.
\label{eq:grant-type}
\end{equation}

These checks ensure that each extracted permission refers to an appropriate resource before it is used for sandbox configuration synthesis.

\subsection{Permission Extraction and Validation}

\sys{} uses an LLM to translate operational requests in repository rules into \dsl{} permission declarations. For example, a request to read a file is represented as an \src{fs.read} permission, while a request to contact an external service is represented as an \src{net.connect} permission. A single instruction may produce multiple declarations when it involves several operations, as in the upload example.
The generated declarations are checked for syntax and type correctness. \sys{} also checks whether each extracted permission is supported by the corresponding instruction in the original rule and whether any requested operations have been omitted. When a problem is identified, the translation is revised before the permissions are used to construct sandbox configurations.
\section{Sandbox Configuration Synthesis}
\label{sec:synthesis}

Permission declarations describe requested access, but a coding agent needs an executable environment in which that access is enforced. \sys{} therefore synthesizes sandbox configurations from the checked declarations and task environment. Synthesis resolves resource names, accounts for dependencies, and determines the access permitted by each configuration. This is necessary because deleting a declaration need not remove exactly the access it names. Executing a script can require its interpreter, while a directory permission can preserve access to a file whose individual permission has been deleted. The full and restricted configurations must account for both relationships before their executions can be compared.

\subsection{Fixing resources and baseline access}

The resource mapping $\beta$ gives declarations a concrete meaning. In the running example, it resolves the workspace path to \src{artifact.tar}, the tool resource to \src{scp}, and the endpoint to \src{victim-host}. These meanings are fixed before constructing the configurations. Removing the archive-read declaration changes whether the sandbox permits reading the archive; it does not substitute a different archive or remove the file from the initial project. Similarly, testing the connection permission retains the same transfer tool and destination. Otherwise, a successful restricted execution could reflect a change in the supplied environment instead of the removal of access.

The environment also supplies baseline permissions $B$ for the task and execution setup. These permissions remain present in every configuration. For the refactoring example, baseline project access allows the agent to read and modify \src{cart.py}, while the rule adds the archive, transfer-tool, and destination requests. A declaration that repeats baseline access does not introduce additional authority. The baseline is fixed independently of the inspected rule, so calling an upload a prerequisite cannot make its permission part of the baseline. Project contents, task request, resource bindings, tools, and execution settings likewise remain fixed across the comparison.

Resolving resources also establishes whether a requested operation has an executable interpretation in the selected environment. A valid resource type does not supply a missing tool or determine an unspecified destination. If the environment cannot interpret a declaration, its removal cannot support a comparison against a full execution containing that request. This distinction keeps the meaning of an unsupported request separate from the outcome of a task test.

\subsection{Interpreting declarations and dependencies}

We compare configurations by the operations they permit, rather than by the number of permission declarations they contain. Write $\Allow(C)$ for the set of allowed operations under configuration $C$. An operation identifies both an action and its resource, such as reading the bound archive or connecting to the bound destination. This interpretation makes overlaps explicit: a read permission on a directory and a read permission on one file beneath it can authorize the same operation. Their declaration identities remain distinct because either may be selected for removal.

For a declaration $p$, write $[\![p]\!]_\beta$ for its interpreted access under the fixed resource mapping and backend, including access needed to realize the requested operation. For example, interpreting execution of a script includes its supported interpreter dependency. Here, $B$ denotes the baseline's allowed operations, including their required support. The full configuration is specified by
\begin{equation}
    \Allow(C)=B\cup\bigcup_{p\in P}[\![p]\!]_\beta.
    \label{eq:configuration-access}
\end{equation}
The equation states which access synthesis must preserve when translating declarations into sandbox controls. Shared dependencies occur in more than one interpreted grant but contribute the same operations to the union. Expanding a declaration does not make its dependencies exclusive to that declaration.

\textbf{Expanding required dependencies.}
We specify dependency expansion over the access entries used to represent the bound permissions. An entry may describe a directory scope, so a finite collection of entries need not enumerate every concrete file operation. Let $D$ be the fixed dependency relation for the selected backend: $(a,b)\in D$ means that realizing access entry $a$ requires entry $b$. Assume that the initial entries and their reachable dependencies form a finite set. For initial entries $X$, expansion is defined by
\begin{equation}
\begin{aligned}
    X_0 &= X,\\
    X_{i+1} &= X_i\cup
       \{b\mid a\in X_i\land(a,b)\in D\}.
\end{aligned}
\label{eq:dependency-expansion}
\end{equation}
Expansion stops when two successive sets are equal. Write $\operatorname{cl}_D(X)$ for the resulting set. Resource bindings and $D$ remain fixed across the full and restricted configurations, so the same retained request has the same dependencies in both.

A worklist gives a direct procedure for computing this set. Initially, the retained set and worklist contain $X$. The procedure removes one entry from the worklist, examines its required entries in $D$, and adds each previously unseen requirement to both collections. An entry already retained is not added again. Consequently, shared dependencies are processed once, and a cycle does not cause repeated expansion. The procedure ends when the worklist is empty. This construction specifies which dependencies synthesis must account for without requiring a particular backend policy format.

\textbf{Proposition (dependency expansion).}
For finite reachable entries and fixed $D$, the worklist terminates and returns the unique least dependency-closed set containing $X$. Moreover,
\begin{equation}
    \operatorname{cl}_D(X\cup X')
    =\operatorname{cl}_D(X)\cup\operatorname{cl}_D(X').
    \label{eq:dependency-union}
\end{equation}
\emph{Proof.}
Each reachable entry enters the worklist at most once, so processing terminates. At termination, every dependency of a retained entry has been retained, establishing closure. Any closed superset of $X$ must contain every added entry, by induction over the additions; this establishes leastness and uniqueness. Every dependency path starting in $X\cup X'$ starts in either $X$ or $X'$, which proves the union identity.

This expansion supplies the dependencies included in $[\![p]\!]_\beta$. The backend interprets the expanded entries as allowed operations and combines them with the fixed baseline. Equation~\ref{eq:dependency-union} explains why shared access remains available after independent deletion. If two declarations require the same interpreter, their expansions overlap on its entry. Removing one declaration leaves that entry in the other's expansion. If the removed declaration is its only source and the baseline does not provide it, the interpreter entry disappears with that declaration. Synthesis therefore recomputes the entries from retained declarations instead of simply removing an entry from the full policy.

The leastness result concerns required dependencies under the chosen bindings and backend. It does not determine which requested operations the coding task needs. That remains the purpose of execution and functional testing. Connecting this construction to $\Allow(C)$ also requires the backend to preserve the operations represented by its expanded entries when installing the sandbox policy.

The output is a configuration for the existing execution environment. It specifies filesystem access, executable tools, and permitted external interactions; it does not replace the task or rewrite the rule to discourage an operation. In particular, the restricted branch still presents the upload instruction to the agent, even when access to its destination is absent. This preserves the question posed by the Introduction: can the agent complete the requested refactoring when the environment prevents the additional operation? Removing the instruction itself would test a different input and would not establish the effect of withholding its permission.

\subsection{Constructing independent restrictions}

For each $p\in P$, \sys{} synthesizes $C^{-p}$ from the original set with only $p$ omitted. The restricted configuration uses the same $\beta$ and $B$ as the full configuration, giving
\begin{equation}
    \Allow(C^{-p})=B\cup
       \bigcup_{p'\in P\setminus\{p\}}[\![p']\!]_\beta.
    \label{eq:restricted-access}
\end{equation}
Every restriction starts from $P$. Testing $p_2$ after $p_1$ therefore does not remove both declarations: it constructs a fresh configuration from $P\setminus\{p_2\}$. This preserves the meaning of each comparison as a test of one requested permission against the same full configuration. It also keeps the baseline available when a requested operation shares resources with ordinary task work.

\textbf{Proposition (access removed by one declaration).}
Assume the full and restricted configurations realize Equations~\ref{eq:configuration-access} and~\ref{eq:restricted-access} under the same bindings and dependency interpretations. Then
\begin{equation}
\begin{aligned}
    \Allow(C^{-p})&\subseteq\Allow(C),\\
    \Allow(C)\setminus\Allow(C^{-p})
       &=[\![p]\!]_\beta\setminus\Allow(C^{-p}).
\end{aligned}
\label{eq:removed-access}
\end{equation}
\emph{Proof.} Each set contributing to $\Allow(C^{-p})$ also contributes to $\Allow(C)$. Moreover, the full union equals $\Allow(C^{-p})\cup[\![p]\!]_\beta$. Subtracting $\Allow(C^{-p})$ gives the second equality.

The proposition isolates what the experiment changes. Removed access comes from the omitted declaration, while operations supported by a remaining declaration or the baseline stay available. If a tree-read grant is removed but a file-read grant remains, that file stays readable. If a tool's interpreter is shared, interpreter access remains. Thus, one deleted declaration can remove only part of its interpreted access. The execution test concerns the access actually removed; it does not require eliminating every way to perform an equivalent computation.

\textbf{Example of overlapping requests.}
Consider a variant of the running example in which $P$ also contains $p_4$, a request to read the workspace directory tree. Permission $p_1$ still requests only the archive read. Since the archive lies beneath that directory, $[\![p_1]\!]_\beta\subseteq[\![p_4]\!]_\beta$. Deleting $p_1$ leaves the broader request in place, so
\begin{equation}
    \Allow(C^{-p_1})=\Allow(C).
    \label{eq:covered-example}
\end{equation}
The restricted agent can still read the archive. A successful refactoring under this configuration therefore says nothing about whether archive access was necessary. Testing $p_4$ asks a different question. The archive remains readable through $p_1$, but reads of other workspace files can be removed if neither the baseline nor another declaration covers them. Equation~\ref{eq:removed-access} identifies precisely that difference, rather than treating the entire directory as inaccessible.

This example also explains why declaration identities must survive policy simplification. The full policy may need only one directory-read entry to represent both requests, but the detector still has two distinct declarations to test. Removing that single entry without resynthesizing the remaining requests would incorrectly remove the archive access supplied by $p_1$. Independent synthesis retains the appropriate declaration after each deletion and then checks which operations actually disappear. The test remains about the original requests even when their compiled policy entries overlap.

\subsection{Checking that access is reduced}

Before interpreting test outcomes, \sys{} compares the full and restricted configurations to check whether allowed access has decreased. A usable restriction requires
\begin{equation}
    \Allow(C^{-p})\subsetneq\Allow(C).
    \label{eq:effective-restriction}
\end{equation}
Together with Equation~\ref{eq:removed-access}, this condition excludes a declaration whose entire interpretation is covered by the remaining configuration. For example, removing an additional read permission on \src{cart.py} changes nothing when baseline access already allows that read. Even if both executions pass every functional test, this deletion cannot show that reading the file was unnecessary: neither execution was prevented from doing so.

Coverage must consider both the action and its resource scope. Permissions on the same file can still differ if one permits reading and the other permits writing. Conversely, different resource names can denote the same file after binding. Synthesis compares their interpreted access, including any access introduced by the selected backend, rather than relying on names alone. For network permissions, the same principle applies to the bound endpoint: a surviving broader grant can cover the destination even after its individual declaration is removed.

In the upload example, deleting $p_3$ must withdraw access to \src{victim-host} while preserving access supplied by $p_1$, $p_2$, and the baseline. If another grant still permits that destination, the deletion is ineffective. Otherwise, the resulting configurations instantiate two sandboxes in which the agent receives the unchanged rule and task, and permission-minimality testing compares their artifacts. Faithful sandbox enforcement is required to connect the configuration comparison to execution: operations excluded by a configuration must remain unavailable to the agent. The set argument alone does not prove enforcement. Nor does synthesis determine which permissions the task needs; that question is tested by the executions in Section~\ref{sec:method}.

\section{Permission-Minimality Testing}
\label{sec:method}

\sys{} tests a permission by comparing two executions of the same task: one with all requested permissions and one with that permission removed. The inputs are the repository rule $R$, initial project $W$, task request $q$, functional tests $T$, and declarations $P$ extracted from $R$. Section~\ref{sec:synthesis} explains how to synthesize the full configuration $C$ from $P$ and a restricted configuration $C^{-p}$ from $P\setminus\{p\}$ for each $p\in P$. This section explains how execution and functional evaluation determine whether the removed permission is unnecessary. The comparison concerns the access that the sandbox actually permits, rather than the presence of a declaration alone.

\subsection{Comparing Full and Restricted Executions}

For each rule, \sys{} first executes the agent under the full configuration $C$, which contains all permissions requested by the rule, and produces an artifact $Y$. For each permission $p$, \sys{} then executes the same rule and task under the corresponding restricted configuration $C^{-p}$, producing an artifact $Y^{-p}$. Importantly, the restricted execution still receives the original rule, including the instruction that requests $p$; only the corresponding permission is removed from the sandbox. Removing the instruction itself would change both what the agent is asked to do and what it is allowed to do, making it difficult to determine whether the permission is actually necessary.

To make the two executions comparable, each execution starts from the same initial project $W$ and uses the same task, tool interfaces, model settings, and execution budget. In particular, the restricted execution starts from the original project rather than the artifact produced by the full execution, so that work performed in one execution cannot carry over to another. The agent is free to react differently when an operation is unavailable. For example, it may observe that an operation is denied, change its plan, and continue the coding task using the remaining permissions. Therefore, the full and restricted executions do not need to follow the same sequence of actions or produce identical patches.

After each execution, \sys{} evaluates the resulting artifact using the same functional tests $T$. We write $\Pass_T(Y)$ when artifact $Y$ passes these tests. The comparison focuses on the final task outcome rather than whether every intermediate tool call succeeds. A denied operation in the restricted sandbox does not necessarily indicate task failure if the agent can still produce a correct artifact. For the running example in Section~\ref{sec:example}, the tests verify both that \src{calculateTotal} is renamed to \src{calculate\_total} and that its original behavior is preserved. The full and restricted artifacts may differ internally as long as both satisfy these requirements.

\subsection{Identifying an Unnecessary Permission}

Passing the functional tests after removing a permission is meaningful only when the restricted configuration actually removes access. As discussed in Section~\ref{sec:synthesis}, permissions may overlap. For example, removing permission to read one file may have no effect if another permission still allows access to the directory containing that file. \sys{} therefore compares the access provided by the full and restricted configurations and requires the restricted configuration to provide strictly less access.

Formally, let $\Allow(C)$ denote the access allowed by configuration $C$. \sys{} identifies permission $p$ as unnecessary for the tested task when both the full and restricted executions pass the functional tests and removing $p$ strictly reduces the available access:

\begin{equation}
\operatorname{Unnecessary}(p)
\iff
\Pass_T(Y)
\land
\Pass_T(Y^{-p})
\land
\Allow(C^{-p}) \subsetneq \Allow(C).
\label{eq:permission-minimality}
\end{equation}

If $\Allow(C^{-p})=\Allow(C)$, removing $p$ has no effective impact and therefore cannot show whether that permission is necessary. Otherwise, when access is reduced but the agent can still complete the task, \sys{} identifies $p$ as unnecessary for the tested task. If at least one requested permission is identified as unnecessary, the corresponding repository rule is flagged as suspicious.

\begin{equation}
\operatorname{Unnecessary}(p)
\iff
\Pass_T(Y)
\land
\Pass_T(Y^{-p})
\land
\Allow(C^{-p}) \subsetneq \Allow(C).
\label{eq:permission-minimality}
\end{equation}

If removing $p$ does not change the available access, for example because another permission still provides the same access, the comparison cannot establish whether $p$ is necessary.

We combine these conditions in the permission-minimality criterion:
\begin{equation}
\begin{aligned}
\operatorname{Unnecessary}(p)\iff{}&
    \Pass_T(Y)\land\Pass_T(Y^{-p})\\
    &\land\Allow(C^{-p})\subsetneq\Allow(C).
\end{aligned}
\label{eq:permission-minimality}
\end{equation}
The access removed by this comparison is $\Allow(C)\setminus\Allow(C^{-p})$. It may be smaller than the access originally associated with $p$ because baseline permissions or other declarations can cover part of the same resource. Thus, identifying $p$ as unnecessary means that the tested task completes without the access actually removed through its deletion. It does not imply that every operation mentioned in $p$ was unavailable. This distinction connects the declaration-level test to the restrictions enforced during execution.

\textbf{Proposition (completion without removed access).}
Assume that the sandbox enforces $\Allow(C^{-p})$ throughout the restricted execution and that $T$ independently and validly checks the task requirements it specifies. If $\operatorname{Unnecessary}(p)$ holds, the restricted execution produces an artifact satisfying those requirements without exercising access in $\Allow(C)\setminus\Allow(C^{-p})$.
\emph{Proof.} Enforcement limits every permitted operation in the restricted execution to $\Allow(C^{-p})$. This set is disjoint from $\Allow(C)\setminus\Allow(C^{-p})$, so the execution cannot exercise any removed access. By Equation~\ref{eq:permission-minimality}, $\Pass_T(Y^{-p})$ also holds. Valid independent evaluation therefore establishes that the resulting artifact satisfies the requirements checked by $T$. These two facts establish the claim.

The proposition makes the role of functional tests explicit. Sandbox enforcement establishes which access was unavailable; the tests establish that the resulting artifact completes the specified task. Neither replaces the other. A passing artifact alone says nothing about the agent's access, while successful confinement alone says nothing about whether the task was completed. The conclusion remains relative to $T$: if the supplied tests omit a legitimate requirement, the comparison does not establish that the removed permission is unnecessary for that omitted requirement.

\subsection{Independent permission tests}
Algorithm~\ref{alg:permission-minimality} applies the criterion to every declaration. Here $\Compile$ denotes configuration synthesis, $\operatorname{Restore}(W)$ creates the initial project state for one execution, and $\Run$ executes the agent with the specified configuration, rule, project, and task. The tool interfaces and execution budget are fixed across all calls to $\Run$. The algorithm returns the set $\mathcal U$ of unnecessary permissions and flags the rule as suspicious when this set is nonempty.

\begin{algorithm}[t]
\caption{Permission-minimality testing}
\label{alg:permission-minimality}
\begin{algorithmic}[1]
\Require Rule $R$; project $W$; task $q$; tests $T$; declarations $P$
\Ensure Unnecessary permissions $\mathcal U$ and suspicious-rule decision
\State $C\gets\Compile(P)$
\State $W_0\gets\operatorname{Restore}(W)$
\State $Y\gets\Run(C,R,W_0,q)$
\State $b\gets\Pass_T(Y)$
\State $\mathcal U\gets\varnothing$
\For{$p\in P$}
    \State $C^{-p}\gets\Compile(P\setminus\{p\})$
    \State $W_p\gets\operatorname{Restore}(W)$
    \State $Y^{-p}\gets\Run(C^{-p},R,W_p,q)$
    \State $b^{-p}\gets\Pass_T(Y^{-p})$
    \If{$\Allow(C^{-p})\subsetneq\Allow(C)$}
        \If{$b\land b^{-p}$}
            \State $\mathcal U\gets\mathcal U\cup\{p\}$
        \EndIf
    \EndIf
\EndFor
\State \Return $(\mathcal U,\mathcal U\neq\varnothing)$
\end{algorithmic}
\end{algorithm}

For $n=|P|$ declarations with runnable configurations, the procedure performs $n+1$ executions: one full execution and one execution per independent deletion. Each restricted configuration is synthesized from the original $P$; deletions never accumulate. The full artifact supplies a common reference for all comparisons. This schedule tests the necessity of each requested permission under otherwise unchanged permissions, without searching through every combination. It also keeps the meaning of each result direct: the reported permission is the declaration removed for that particular comparison.

In the running example, testing $p_3$ removes access to \src{victim-host} while preserving the rule's upload instruction. The agent may receive a connection denial and still complete the rename. If the full and restricted artifacts both pass $T$, and no remaining permission covers the destination, $p_3$ satisfies Equation~\ref{eq:permission-minimality}. The rule is consequently flagged as suspicious. An upload attempt is not required: the restriction establishes that the removed access was unavailable regardless of whether the agent tried to use it. Permissions $p_1$ and $p_2$ are tested separately against the same full configuration.

Independent deletion does not establish that all unnecessary permissions can be removed together. Suppose a rule requests two distinct editing tools that can each perform the same function rename. With access to both tools, the task succeeds. Removing either tool separately can also succeed because the other remains available. Removing both may prevent the task from being completed. Each individual conclusion is still valid for its tested configuration, but combining the conclusions would require another execution. \sys{} therefore identifies individually unnecessary permissions, rather than constructing a globally minimum permission set.

A failed restricted execution does not prove that its removed permission is necessary. The agent may choose an unsuccessful plan, exhaust its budget, or fail for reasons unrelated to the restriction. Similarly, when the full execution fails, the pair cannot satisfy Equation~\ref{eq:permission-minimality}. The method relies on successful comparisons to identify unnecessary access; it does not interpret the absence of such a comparison as proof of necessity. A suspicious-rule decision likewise identifies a request that exceeds the tested task's needs, not proof of malicious intent: legitimate optional workflow operations may also be unnecessary for that task.

\section{System Implementation}
\label{sec:implementation}

\sys{} supports sandbox execution on both Linux and macOS. Each sandbox isolates the coding agent from the host system and enforces the filesystem, tool, and network permissions specified by the synthesized configuration. On macOS, \sys{} can either use the native sandbox mechanism or run the Linux backend through Docker. Before each execution, \sys{} restores the project to the same initial state so that the full and restricted executions start from the same codebase. We implement the executor as a custom tool-calling coding agent accessed through OpenRouter, and use a Python-based test adapter to evaluate the generated artifacts.
\section{Evaluation Settings}
\label{sec:evaluation}

\subsection{Research Questions and Metrics}

We evaluate \sys{} through the following two research questions:

\begin{itemize}[leftmargin=*]
    \item \textbf{RQ1: How effective is \sys{} in detecting malicious repository rules?}
    We measure the detection rate on attack rules and the false-positive rate on benign rules.

    \item \textbf{RQ2: What is the runtime and cost of \sys{}?}
    We measure the average end-to-end runtime and API cost per rule, including permission translation and sandbox execution.
\end{itemize}

\subsection{Datasets and Experimental Setup}

We evaluate \sys{} on 314 attack rules from AIShellJack~\cite{aishelljack} and 85 benign rules, including five benign rules from AIShellJack and 80 manually verified benign rules from GHAgentFiles~\cite{ghagentfiles}. The AIShellJack attack rules contain prompt-injection payloads, while the GHAgentFiles rules are selected from real-world repository instruction files.
All rules are evaluated using the same Python task, which asks the agent to rename \src{calculateTotal} to \src{calculate\_total} while preserving its behavior and the \src{discountRate} parameter. The functional tests check the renamed interface, removal of the old function name, and the expected outputs for empty, regular, and discounted inputs.
We use \src{openai/gpt-5.5} through OpenRouter for both permission translation and agent execution, with seed~0, low reasoning effort, and the provider-default temperature. Translation and agent responses are limited to 8,192 and 4,096 tokens, respectively. Each execution is limited to eight agent turns with a ten-second command timeout. Experiments run with four workers in a fixed Linux Docker environment on an Apple M4 Pro with 24\,GiB of memory and macOS~26.6.2.
\section{Evaluation Results}
\label{sec:results}

\subsection{RQ1: Effectiveness}

Table~\ref{tab:detection-results} summarizes the detection results of \sys{}. \sys{} successfully detects all 314 attack rules, achieving a detection rate of 100\%. Among the 85 benign rules, only three are incorrectly flagged as suspicious, corresponding to a false-positive rate of 3.5\%. The remaining 82 benign rules raise no alarm.
These results show that permission-minimality testing identifies attack rules in this evaluation even when the requested refactoring can be completed. Unnecessary access is task-relative: a permission that is unnecessary for this refactoring may support a legitimate development workflow. The three false positives therefore motivate evaluating rules against tasks that cover their intended use, rather than interpreting unnecessary access as proof of malicious intent.

\begin{table}[t]
\caption{Detection results of \sys{} on attack and benign repository rules.}
\label{tab:detection-results}
\centering\small
\begin{tabular}{@{}lrrr@{}}
\toprule
Population & Inputs & Detected & Not detected \\
\midrule
Attack rules & 314 & 314 & 0 \\
Benign rules & 85 & 3 & 82 \\
\bottomrule
\end{tabular}
\end{table}

\finding{RQ1: Effectiveness}{\sys{} detects all 314 attack rules (100\%) and reports only three false positives among the 85 benign rules (3.5\%).}

\subsection{RQ2: Runtime and Cost}

We measure the end-to-end runtime and API cost of \sys{} on all 85 benign rules. On average, permission translation and sandbox execution take 44.35 seconds and cost approximately US\$0.22 per rule. The runtime covers the complete testing process, including permission translation, sandbox construction, agent execution, and functional testing.
For a rule containing $n$ requested permissions, \sys{} shares one full execution across all comparisons and performs one restricted execution per permission, requiring $n+1$ executions rather than $2n$. Rules with more extracted permissions may consequently cost more to evaluate, even when their source files are short. Per-rule cost also depends on the number of agent turns and the work performed in each execution.

\finding{RQ2: Runtime and Cost}{Across all 85 benign rules, \sys{} takes an average of 44.35 seconds and costs approximately US\$0.22 per rule.}

\section{Threats to Validity}
\label{sec:threats}

A potential threat to internal validity is the randomness of LLM-based permission translation and agent execution, which may lead to different results across runs. To mitigate this threat, we use the same model configuration, task, and initial project state for the full and restricted executions. A potential threat to external validity is that our evaluation uses 314 attack rules, 85 benign rules, one coding task, and one agent model, so the results may not generalize to other rules, tasks, or agents. To mitigate this threat, we include both attack rules and real-world benign rules and leave broader evaluation across more tasks and models to future work. A potential threat to construct validity is that a permission unnecessary for the tested task may still be useful for another development task, while functional tests may not capture every aspect of task correctness. To mitigate this threat, we evaluate all rules under the same clearly specified task and functional tests and interpret the reported results with respect to this evaluation setting.
\section{Related Work}
\label{sec:related}

\subsection{Prompt Injection and Agent Security}

Indirect prompt injection occurs when an LLM treats attacker-controlled content as instructions~\cite{indirectpi}. Existing defenses address this problem from different perspectives. Instruction Hierarchy, StruQ, and SecAlign improve the separation between trusted instructions and untrusted content through training or structured inputs~\cite{hierarchy,struq,secalign}, while DataSentinel and PIGuard detect injected instructions~\cite{datasentinel,piguard}. Task Shield determines whether instructions and tool calls are aligned with the user's intended task~\cite{taskshield}. Progent and DRIFT regulate agent tool use through policies or execution plans~\cite{progent,drift}, while CaMeL and Fides restrict information propagation during agent execution~\cite{camel,fides}.

The closest work to \sys{} is SkillScope, which studies excessive privileges through task-conditioned counterfactual execution~\cite{skillscope}. SkillScope removes or neutralizes candidate actions and checks whether the task can still proceed. In contrast, \sys{} keeps the repository rule unchanged, removes one requested permission through sandbox configuration, and directly checks whether the resulting software artifact still passes the functional tests. Our work therefore focuses on testing whether permissions requested by coding-agent rules are necessary for completing a concrete software development task.

\subsection{Permission Control and Sandboxing}

Prior work has extensively studied how software permissions can be derived and enforced. Sandbox mining infers policies from accesses observed during execution, including work on Android applications and Linux containers~\cite{miningsandboxes,mininglinux}. Capsicum provides capability-based isolation, while Shill and capweave enforce or synthesize capability policies for program execution~\cite{capsicum,shill,capweave}. Abhaya synthesizes system-call policies using static analysis, and DynBox restricts system calls across different execution phases~\cite{abhaya,dynbox}.

These approaches primarily aim to construct or enforce an appropriate sandbox policy. \sys{} addresses a different question: given permissions explicitly requested by a repository rule, are those permissions actually necessary for completing the specified coding task? To answer this question, \sys{} constructs a full sandbox and restricted sandboxes in which individual permissions are removed, and compares task completion across these executions. Related systems such as DeltaBox and YoloFS provide mechanisms for isolated state restoration and filesystem control~\cite{deltabox,yolofs}, which are complementary to our permission-minimality testing approach.

\subsection{Coding-Agent Rules and Security Benchmarks}

Recent work has studied both repository-level instructions and security threats to coding agents. AIShellJack demonstrates prompt-injection attacks through poisoned coding resources~\cite{aishelljack}, while InjecAgent and AgentDojo evaluate indirect prompt injection through malicious tool responses and attacker objectives~\cite{injecagent,agentdojo}. These studies provide benchmarks for evaluating whether agents follow malicious instructions.

Repository-level context has also received increasing attention. GHAgentFiles collects large-scale histories of coding-agent instruction files, while CTXbench studies the usefulness of repository context for coding agents~\cite{ghagentfiles,ctxbench}. SWE-bench provides real-world repository tasks together with executable tests for evaluating whether coding agents successfully complete software engineering tasks~\cite{swebench}. Building on these directions, our evaluation combines attack rules with manually verified benign repository rules and tests whether permissions requested by these rules are necessary for completing a specified coding task.
\section{Conclusion and Future Work}
\label{sec:conclusion}

We present \sys{}, a permission-minimality testing approach for detecting suspicious repository rules in coding agents. \sys{} translates repository rules into permission declarations, constructs full and restricted sandboxes by removing one permission at a time, and tests whether the coding task can still be completed under the restricted permissions. Our evaluation shows that \sys{} detects all 314 attack rules (100\%) while reporting only three false positives among 85 benign rules (3.5\%). On average, permission translation and sandbox execution take 44.35 seconds and cost approximately US\$0.22 per rule.

Future work will evaluate \sys{} on more diverse repository rules, coding tasks, and agent models to further assess its generalizability. We also plan to compare \sys{} with additional prompt-injection and task-alignment defenses under the same evaluation settings and further improve permission translation and sandbox enforcement.


\bibliographystyle{ACM-Reference-Format}
\bibliography{references}

\end{document}